\documentstyle[art12,makeidx,bezier]{article}
\newcommand{\lb}[1]{\label{eqn:#1}}
\newcommand{\rf}[1]{\ref{eqn:#1}}
\newcommand{\bm}[1]{\mbox{\boldmath$#1$}}
\begin{document}
\baselineskip 20pt
\vglue 2cm
\centerline{{\Large{\bf A Realization of Discrete Geometry}}}
\vglue 0.5cm
\centerline{{\Large{\bf by String Model}}}

\vglue 3cm
\centerline{Satoru SAITO\footnote{e-mail : saito@phys.metro-u.ac.jp}}
\vglue 2cm
\centerline{{\it Department of Physics}}
\centerline{{\it Tokyo Metropolitan University}}
\centerline{{\it Minamiohsawa 1-1, Hachiohji,}}
\centerline{{\it  Tokyo 192-0397, Japan}} 
\vglue 6cm
\noindent
{\bf Abstract}

A realization of discrete conjugate net is presented by using correlation functions of strings in a gauge covariant form.

\vfill\eject
\section{Introduction}

The recent study of integrable systems has revealed unexpected relationship among different subjects in theoretical physics. Various integrable systems are often combined into a single discrete integrable equation. Such a discrete equation can be interpreted differently depending on the subject. For example the Hirota bilinear difference equation (HBDE) includes all of soliton equations which belong to the KP hierarchy in various continuum limits\cite{Hirota,Miwa}. The same equation is satisfied by string correlation functions in particle physics\cite{S} as well as by transfer matrices of solvable lattice models in statistical physics\cite{KLWZ}. 

From mathematical point of view HBDE is nothing but Fay's identity\cite{Fay} which characterizes algebraic curves. On the other hand some of soliton equations, such as sine-Gordon equation, Toda lattice etc., had been classical subjects studied in differential geometry or projective geometry. Therefore it is quite natural that a geometry of discrete surface has been recently developed\cite{BP,BK,DS} and discrete version of soliton equations, including HBDE itself, appear as equations satisfied by curvature of the discrete surface. This fact means that the deformation of integrable systems to discrete space preserves integrability when the deformation satisfies certain geometrical constraints.

There have been proposed several extentions of discrete integrable systems to higher dimensional spaces\cite{DS}-\cite{DMAMS}. We do not know at this moment if there exists some correlation among different extensions. It is, therefore, desirable to understand their mathematical backgrounds. Physical interpretation of the extensions is another problem to be studied. A beautiful interpretation by means of solvable lattice models has been given\cite{LWZ} to the higher dimensional extension of HBDE. Some physical implication of tetrahedron equation has been also discussed\cite{KKS}. The purpose of this report is to show that the higher dimensional extension of the discrete conjugate net\cite{DS,DMAMS} can be naturally interpreted by means of the string theory.

Brief reviews of the discrete conjugate net and the string theory will be given in \S 2 and \S 3. We present in \S 4 a realization of the discrete conjugate net in terms of correlation functions of strings. 
\vfill\eject
\section{Conjugate nets on a lattice}

The concept of conjugate nets on a lattice space has been discussed by Doliwa and Santini\cite{DS,DMAMS}. Let us summarize briefly, throughout this section, some of their results, which are necessary in our discussion.

\noindent
\underline{Laplace equation of quadrilateral lattice}:

Let $\bm{q}=(q_1,q_2,\cdots,q_D)\in \bm{Z}^D$ be the $D$ dimensional lattice space. Then the $D$-dim quadrilateral lattice \bm{x} is a map from the lattice \bm{q} to $\bm{R}^M,\quad M\ge D$, such that every elementary quadrilateral is planar. It can be characterized by the Laplace equation,
\begin{equation}
\Delta_\mu\Delta_\nu\bm{x}-T_\mu\left((\Delta_\nu H_\mu)H_\mu^{-1}\right)\Delta_\mu\bm{x}-T_\nu\left((\Delta_\mu H_\nu)H_\nu^{-1}\right)\Delta_\nu\bm{x}=0,
\lb{discrete Laplace}
\end{equation}
where $T_\mu$ is the shift operator which brings $q_\mu$ into $q_\mu+1$ and $\Delta_\mu:=T_\mu-1$. The Lame coefficient $H_\rho$ itself satisfies
\begin{equation}
\Delta_\mu\Delta_\nu H_\rho-T_\mu\left((\Delta_\nu H_\mu)H_\mu^{-1}\right)\Delta_\mu H_\rho-T_\nu\left((\Delta_\mu H_\nu)H_\nu^{-1}\right)\Delta_\nu H_\rho=0.
\lb{discrete Darb}
\end{equation}

If we define the tangent vector ${\bm{X}}_\nu$ by
$$
\Delta_\nu\bm{x}=(T_\nu H_\nu)\bm{X}_\nu,\qquad \nu=1,2,\cdots, D,
$$
then $(\rf{discrete Laplace})$ and $(\rf{discrete Darb})$ turn out to be
\begin{equation}
\Delta_\mu\bm{X}_\nu=(T_\mu Q_{\nu\mu})\bm{X}_\mu,\qquad \mu\ne \nu,
\lb{Discrete Laplace}
\end{equation}
\begin{equation}
\Delta_\mu H_\nu=(T_\mu H_\mu)Q_{\mu\nu},\qquad \mu\ne \nu.
\lb{Discrete Lame}
\end{equation}
The consistency relation of these equations yields for the rotation coefficient $Q_{\mu\nu}$,
\begin{equation}
\Delta_\rho Q_{\mu\nu}=(T_\rho Q_{\mu\rho})Q_{\rho\nu},\qquad \mu\ne \nu\ne \rho\ne \mu.
\lb{Discrete Darboux}
\end{equation}

Equations $(\rf{Discrete Laplace})$, $(\rf{Discrete Lame})$ and $(\rf{Discrete Darboux})$ are discrete analogue of the following classical results for the continuous conjugate nets parameterized by $u_\mu,\ \mu=1,2,\cdots, D$ :
\begin{equation}
{\partial \bm{X}_{\mu}\over \partial u_\nu}=\beta_{\mu\nu}\bm{X}_\nu,\qquad \mu\ne \nu,
\lb{tangent vector eq}
\end{equation}
\begin{equation}
{\partial H_{\mu}\over \partial u_\nu}=\beta_{\nu\mu}H_\nu,\qquad \mu\ne \nu,
\end{equation}
\begin{equation}
{\partial \beta_{\mu\nu}\over \partial u_\rho}=\beta_{\mu\rho}\beta_{\rho\nu},\qquad \mu\ne \nu\ne \rho\ne \mu.
\lb{Darboux}
\end{equation}

\noindent
\underline{Focal lattice and Laplace transform} : 

The $D$-dim rectilinear congruence $\bm{l}(\bm{q})$ is a map from the lattice to lines in $\bm{R}^M$, such that every two neighbouring lines is coplanar. If all points of the quadrilateral lattice $\bm{x}(\bm{q})$ belong to the rectilinear congruence \bm{l}(\bm{q}), \bm{x} and \bm{l} are called conjugate.

The focal lattice $\bm{y}_\mu(\bm{l})$ is a lattice constructed out of the intersection points of the lines $\bm{l}(\bm{q})$ and $\bm{l}(\bm{q}+\bm{e}_\mu).$ Then the following theorem was shown.

\noindent
{\bf Theorem}\ :\ {\it Focal lattices of congruences conjugate to quadrilateral lattices are quadrilateral lattices.}

The map from \bm{x} to the forcal lattice $\bm{y}_\mu(\bm{l})$ is called Laplace transformation and is given by
\begin{equation}
{\cal L}_{\mu\nu}(\bm{x})=\bm{y}_\nu(l_\mu(\bm{x}))=\bm{x}-{H_\mu\over Q_{\mu\nu}}\bm{X}_\mu.
\end{equation}
Doliwa has shown\cite{DS} that HBDE arises as an equation satisfied by invariants under this transformation. On the other hand the map of \bm{x} to the quadrilateral lattice conjugate to the $\mu$-th tangent congruence of \bm{x} is called Darboux transformation and is given by
\begin{equation}
{\cal L}_\mu(\bm{x})=\bm{x}-{\phi\over\Delta_\mu\phi}\Delta_\mu\bm{x},
\end{equation}
where $\phi$ is a solution of $(\rf{discrete Laplace})$.

\section{Elements of string theory}

Analytical property of the string correlation functions in particle physics is characterized by Hirota bilinear difference equation\cite{S}. In this section I like to review briefly the elements of the string theory. 

Let $z\in \bm{C}$ be a complex proper time of a string. Then the positive and negative oscillation parts of the string coordinate 
\begin{equation}
X^\mu(z):=X^\mu_+(z)+X^\mu_-(z),\quad \mu=1,2,\cdots, D
\end{equation}
satisfy
\begin{equation}
[X^\mu_+(z),\ X^\nu_-(z')]=\delta^{\mu\nu}\ln (z-z').
\end{equation}

Interaction of a ground state particle with the string takes place through the vertex operator  
\begin{equation}
V(\bm{k},z):=\exp\left[i\sum_\mu k_\mu X_+^\mu(z)\right]\exp\left[i\sum_\mu k_\mu X_-^\mu(z)\right].
\lb{vertex}
\end{equation}
where $\bm{k}\in \bm{R}^{D}$ is the momentum of the interacting particle. Since the string is quantized, the vertex operators do not commute but enjoy the following relations:
\begin{eqnarray}
V(\bm{k},z)V(\bm{k}',z')&=&(z-z')^{\bm{k}\bm{k}'}:V(\bm{k},z)V(\bm{k}',z'):\nonumber\\
&=&(-1)^{\bm{k}\bm{k}'}V(\bm{k}',z')V(\bm{k},z).
\end{eqnarray}

For a given configuration of strings, say $|G\rangle$, the correlation function can be calculated
\begin{equation}
F_G(\bm{k}_1,\bm{k}_2,\cdots,\bm{k}_N):=\langle 0|V(\bm{k}_1,z_1)V(\bm{k}_2,z_2)\cdots V(\bm{k}_N,z_N)|G\rangle
\end{equation}
where $|0\rangle$ is the vacuum state annihilated by $X_-^\mu(z)$. When $|G\rangle$ itself is the vacuum we can calculate the amplitude explicitly and obtain 
\begin{eqnarray}
F_0(\bm{k}_1,\bm{k}_2,\cdots,\bm{k}_N)&:=&\langle 0|V(\bm{k}_1,z_1)V(\bm{k}_2,z_2)\cdots V(\bm{k}_N,z_N)|0\rangle\nonumber\\
&=&
\prod_{l>j}(z_j-z_l)^{\bm{k}_j\bm{k}_l}.
\end{eqnarray}

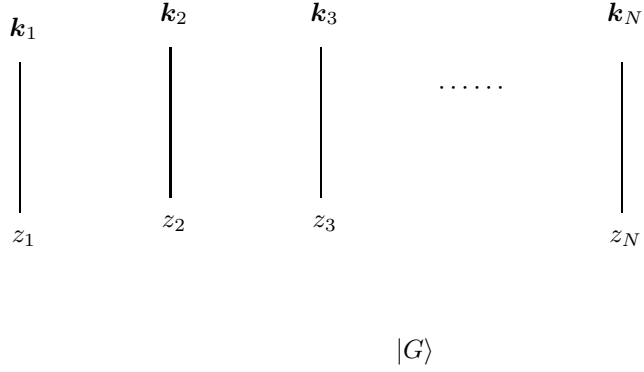
\begin{center}\begin{picture}(100,55)
\put(9,48){\makebox(3,3)[c]{$\bm{k}_1$}}
\put(29,50){\makebox(3,3)[c]{$\bm{k}_2$}}
\put(49,50){\makebox(3,3)[c]{$\bm{k}_3$}}
\put(89,50){\makebox(3,3)[c]{$\bm{k}_N$}}
\put(10,25){\line(0,1){20}}
\put(30,27){\line(0,1){20}}
\put(50,27){\line(0,1){20}}
\put(90,25){\line(0,1){20}}
\put(65,40){\makebox(10,3)[c]{$\cdots\cdots$}}
\put(9,20){\makebox(3,3)[c]{$z_1$}}
\put(29,22){\makebox(3,3)[c]{$z_2$}}
\put(49,22){\makebox(3,3)[c]{$z_3$}}
\put(89,20){\makebox(3,3)[c]{$z_N$}}
\put(60,5){\makebox(5,3)[c]{$|G\rangle$}}
\end{picture}\end{center}

It was shown in \cite{S} that if we define
\begin{equation}
f(\bm{k}_1,\bm{k}_2,\cdots,\bm{k}_N)={F_G(\bm{k}_1,\bm{k}_2,\cdots,\bm{k}_N)\over F_0(\bm{k}_1,\bm{k}_2,\cdots,\bm{k}_N)}
\lb{f=tau/tau}
\end{equation}
it solves the HBDE
$$
z_a(z_b-z_c)f(\bm{k}_a+\bm{e}_\mu,\bm{k}_b,\bm{k}_c)f(\bm{k}_a,\bm{k}_b+\bm{e}_\mu,\bm{k}_c+\bm{e}_\mu)\qquad\qquad\qquad\qquad
$$
$$
\qquad\quad
+ z_b(z_c-z_a)f(\bm{k}_a,\bm{k}_b+\bm{e}_\mu,\bm{k}_c)f(\bm{k}_a+\bm{e}_\mu,\bm{k}_b,\bm{k}_c+\bm{e}_\mu)\qquad\qquad
$$
\begin{equation}
\qquad\qquad\qquad\qquad
+ z_c(z_a-z_b)f(\bm{k}_a,\bm{k}_b,\bm{k}_c+\bm{e}_\mu)f(\bm{k}_a+\bm{e}_\mu,\bm{k}_b+\bm{e}_\mu,\bm{k}_c)=0,
\end{equation}
where $\bm{k}_a,\ \bm{k}_b,\ \bm{k}_c$ are any three of $\bm{k}_1,\bm{k}_2,\cdots,\bm{k}_N$ and $\bm{e}_\mu$ denotes the unit vector along the $\mu$ direction.
\vfill\eject
\section{String realization of the conjugate net}

We have discussed two subjects, discrete geometry and string models, in the previous sections. They belong to different fields. But they share HBDE in common. This fact might imply that the discrete geometry is realized by the string model in particle physics. I like to show that this happens to be correct. The main part of the argument owes to the observation in \cite{DMAMS}\footnote{I like to thank Dr. A.Doliwa for his presentation of this paper prior to its publication.}, in which the KP hierarchy was described in terms of discrete geometry. On the other hand the correspondence between the KP hierarchy and the string model was established in \cite{S}. Therefore the problem which is left for us is to find a way of the translation from one language to another. In conclusion we will see that the string model variables can be associated directly to the discrete conjugate net. Hence the string model interpretation of the discrete conjugate net turns out to be quite natural.

\noindent
\underline{KP correspondence} :

The KP hierarchy\cite{Hirota,Miwa,DJKM} is a set of infinitely many bilinear differential equations of Hirota type. All of their solutions are given by a single function $\tau(t_1, t_2, t_3, \cdots)$, which is called the KP $\tau$ function. $t_n,\ n=1,2,\cdots$ are the soliton variables along which solitons can propagate. The $D$-component KP hierarchy\cite{DJKM} is an extension of the KP hierarchy in which the soliton variables $t_n$'s are generalized to $D$-component vectors $\bm{t}_n$'s. In addition we need charge variables $\bm{q}$ which take values on the $D$-dimensional lattice space. We write the corresponding $\tau$ function as $\tau(\bm{q}, \bm{t}_1, \bm{t}_2, \bm{t}_3, \cdots)$.

Then the matrix function, which is defined by ($\epsilon_{\mu\nu}:=sgn(\nu-\mu),\  \epsilon_{\mu\mu}:=1$)
\begin{equation}
W_{\mu,\nu}(z):=\epsilon_{\mu\nu}z^{\delta_{\mu\nu}-1}{\tau\left(\bm{q}+\bm{e}_\mu-\bm{e}_\nu,\bm{t}_1-z\bm{e}_\nu,\bm{t}_2-{z^2\over 2}\bm{e}_\nu,\bm{t}_3-{z^3\over 3}\bm{e}_\nu,\cdots\right)\over\tau(\bm{q},\bm{t}_1,\bm{t}_2,\bm{t}_3,\cdots)}e^{\sum_nz^{-n}t_{n\nu}},
\lb{W}
\end{equation}
satisfies the linear equation\cite{DJKM}
\begin{equation}
{\partial W_{\mu,\nu}\over \partial t_{n\rho}}=B_{\mu,\rho}^{(n\rho)}W_{\rho,\nu},
\lb{Sato eq}
\end{equation}
where $B_{\mu,\rho}^{(n\rho)}$ is a matrix differential operator.

Doliwa {\it et al} have found a geometrical interpretation of the multicomponent KP hierarchy based on their theory of conjugate net\cite{DMAMS}. According to their argument the soliton variables $\bm{t}_1$ are interpreted as the coordinates \bm{u} of the conjugate net, whereas the rest of variables $\bm{t}_n,\ n\ge 2$ describe iso-conjugate deformations of the nets. Then they associate $W_{\mu,\nu}$ with the componet of the tangent vector $\left(\bm{X}_\mu\right)_\nu$ and $B_{\mu\nu}$ with the rotation matrix $\beta_{\mu\nu}$, so that the correspondence between $(\rf{tangent vector eq})$ and $(\rf{Sato eq})$ follows :
\begin{equation}
\bm{t}_1\ \leftrightarrow\ \bm{u},\quad W_{\mu,\nu}\ \leftrightarrow\ \left(\bm{X}_\mu\right)_\nu,\quad B_{\mu\nu}\ \leftrightarrow\ \beta_{\mu\nu}.
\end{equation}
In terms of the KP $\tau$ function $\beta_{\mu\nu}$ is thus given by:
\begin{equation}
\beta_{\mu\nu}=\epsilon_{\mu\nu}{\tau\left(\bm{q}+\bm{e}_\mu-\bm{e}_\nu,\bm{t}_1,\bm{t}_2,\bm{t}_3,\cdots\right)\over\tau(\bm{q},\bm{t}_1,\bm{t}_2,\bm{t}_3,\cdots)}.
\lb{beta}
\end{equation}
Moreover an operation of the vertex operator of the KP theory to the matrix $W$ is interpreted as a Levy transformation of the tangent vector \bm{X} according to
\begin{equation}
{\cal L}_\mu\left(\bm{X_\nu}\right)\ \leftrightarrow\ 
z^{-\delta_{\mu\nu}}\exp\left[-\sum_{n=1}^\infty {1\over n}z^{n}{\partial\over\partial t_{n\mu}}\right]\bm{W}_\nu,
\lb{L}
\end{equation}
where $\bm{W}_\nu$ is the $\nu$th row vector of the matrix $W$.

\noindent
\underline{String realization} :

Now we will proceed to show the realization of the lattice conjugate nets by means of string theory. The key observation is the Miwa transformation\cite{Miwa}, which enables us to interprete the string theory in terms of the soliton theory\cite{S}. It is given by
\begin{equation}
\bm{t}_0:=\sum_{j=1}^N \bm{k}_j,\qquad \bm{t}_n={1\over n}\sum_{j=1}^N\bm{k}_jz_j^{n},\quad n=1,2,3,\cdots
\lb{Miwa}
\end{equation}
Miwa did not introduce $\bm{t}_0$ in \cite{Miwa}, but it is natural to incorpolate it into the string theory as the center of mass momentum of external particles. 

It was shown\cite{S} that, after the change of variables, the KP $\tau$ function is the same as the ratio of the string correlation functions in $(\rf{f=tau/tau})$,
\begin{equation}
\tau(\bm{t}_0, \bm{t}_1, \bm{t}_2, \bm{t}_3, \cdots)={F_G(\bm{k}_1, \bm{k}_2, \bm{k}_3, \cdots, \bm{k}_N)\over F_0(\bm{k}_1, \bm{k}_2, \bm{k}_3, \cdots, \bm{k}_N)}.
\end{equation}
From this we first notice that the center of mass momentum $\bm{t}_0$ must be interpreted as the fundamental lattice \bm{q} of the conjugate net. The other coordinates are related through the second equation of $(\rf{Miwa})$. It is easy to verify the following identities,
\begin{equation}
{\partial\over\partial k_{j\mu}}={\partial\over\partial t_{0\mu}}+\sum_{n=1}^\infty{1\over n}z_j^n{\partial\over\partial t_{n\mu}},\qquad j=1,2,\cdots, N.
\lb{d/dk}
\end{equation}
We denote by $T_{j\mu}$ the operation which shifts $\bm{k}_j$ to $\bm{k}_j+\bm{e}_\mu$,
\begin{equation}
T_{j\mu}:=\exp\left[{\partial\over\partial k_{j\mu}}\right].
\lb{T_jmu}
\end{equation}
Then we can prove
\begin{equation}
F_0^{-1}T_{j\mu}F_0=\prod_l(z_j-z_l)^{k_l}=z_j^{t_{0\mu}}\exp\left[-\sum_{n=1}^\infty z_j^{-n}t_{n\mu}\right].
\end{equation}
Let us recall that the total momentum $\bm{t}_0$ of the correlation function $F_0$ does not change because it has no background, while one of $F_G$ can be changed by shifting the momentum of the state $|G\rangle$ to the opposite, simultaneously. We denote by $T_{\mu}$ the shift of $\bm{q}$ into $\bm{q}+\bm{e}_\mu$. This is exactly what we defined before. Combining all these together the tangent vector, which was once given by $(\rf{W})$, can be rewritten as
\begin{equation}
\left(\bm{X}_\mu\right)_{j\nu}=\epsilon_{\mu\nu}F_G^{-1}T_\mu T_{j\nu}^{-1}F_G.
\lb{X}
\end{equation}
Similarly the rotation matrix is translated into
\begin{equation}
Q_{\mu\nu}=\epsilon_{\mu\nu}F_G^{-1}T_\mu T_\nu^{-1}F_G.
\lb{Q_munu}
\end{equation}

The operator of the Darboux transformation is proportional to the shift operator $T_{j\mu}^{-1}$ itself, as we see by comparison of $(\rf{L})$, $(\rf{d/dk})$ and $(\rf{T_jmu})$. It will be, however, more convenient to define it according to
\begin{equation}
{\cal L}_{j\mu}:=F_0^{-1}T_{j\mu}^{-1}F_0.
\lb{covariant Darboux}
\end{equation}
In this expression it is assumed that the operator $T_{j\mu}^{-1}$ acts to all of functions on its right. Under this condition it is nothing but the vertex operator $(\rf{vertex})$ of the string theory.

From these correspondences we can draw some picture of the lattice conjugate net which is realized by the string models. Let us consider the space of momenta of $N$ external particles $(\bm{k}_1, \bm{k}_2, \cdots, \bm{k}_N)$. We also consider the momenta of the background, so that the total momenta \bm{q} of the external particles becomes independent variable. Suppose the background space is compactified into torus in all $D$ directions. Then its momenta, hence \bm{q} itself, can take only integer values. We consider the lattice space of $\bm{q}\in \bm{Z}^D$ as the fundamental lattice of the conjugate net which is embedded in the space $\bm{R}^{DN}$ of the momenta of the $N$ external particles.

Based on this setting we can interprete the tangent vector $\bm{X}_\mu$ of $(\rf{X})$ by the gauge covariant shift operator along the $\mu$-direction of the \bm{q} space. Similarly the rotation operator $Q_{\mu\nu}$ of $(\rf{Q_munu})$ has the meaning of a gauge covariant rotation operator in this space. The Darboux transformation operator of $(\rf{covariant Darboux})$ is somewhat different. It acts on the space of momenta of external particles $\bm{k}_j\in \bm{R}^{DN}$, again in a gauge covariant form. These manifestly gauge covariant forms make the string realization of the conjugate net quite natural.

\noindent
{\bf Acknowledgement}

I would like to thank Profs.Y.Nakamura, J.Matsukidaira and other organizers of this workshop for invitation. This work is supported in part by the Grant-in-Aid for General Scientific Research from the Ministry of Education, Science, Sports and Culture, Japan (No 10640278).

\end{document}